\documentclass[superscriptaddress,preprint,floatfix,amsmath,aps]{revtex4-1}
\usepackage{amsmath,amssymb,latexsym,mathrsfs,amsfonts}
\usepackage{graphicx}
\usepackage{subfigure}

\usepackage{dcolumn}% Align table columns on decimal point
\usepackage{epstopdf}
\usepackage{bm}% bold math
\usepackage{flushend}
\usepackage{multirow}
\usepackage{color}
\usepackage[mathlines]{lineno}
\usepackage{soul}
\usepackage{mathtools}
\usepackage[normalem]{ulem}

\newcommand{\ccs}{CsCr$_3$Sb$_5$}
\newcommand{\avs}{$A$V$_3$Sb$_5$}

\newpage
\newpage

\begin{document}

%\title{Frustrated Altermagnetism and Charge Density Wave in Kagome Superconductor CsCr$_3$Sb$_5$} \preprint{1}
\title{Altermagnetic Ground State in Distorted Kagome Metal CsCr$_3$Sb$_5$ }

\author{Chenchao Xu}
 \affiliation{School of Physics, Hangzhou Normal University, Hangzhou 310036, P. R. China}
 \affiliation{Center for Correlated Matter, Zhejiang University, Hangzhou 310058, China}

\author{Siqi Wu}
 \affiliation{School of Physics, Zhejiang University, Hangzhou 310058, China}
 
\author{Guo-Xiang Zhi}
 \affiliation{Tianmushan Laboratory, Hangzhou 310023, China}

\author{Guanghan Cao}
 \affiliation{School of Physics, Zhejiang University, Hangzhou 310058, China}
 \affiliation{Institute for Advanced Study in Physics, Zhejiang University, Hangzhou 310058, China}

\author{Jianhui Dai}
 \affiliation{School of Physics, Hangzhou Normal University, Hangzhou 310036, P. R. China}
 \affiliation{Institute for Advanced Study in Physics, Zhejiang University, Hangzhou 310058, China}

\author{Chao Cao}
 \email[E-mail address: ]{ccao@zju.edu.cn}
 \affiliation{Center for Correlated Matter, Zhejiang University, Hangzhou 310058, China}
 \affiliation{School of Physics, Zhejiang University, Hangzhou 310058, China}
 \affiliation{Institute for Advanced Study in Physics, Zhejiang University, Hangzhou 310058, China}

\author{Xiaoqun Wang}
 \email[E-mail address: ]{xiaoqunwang@zju.edu.cn}
 \affiliation{School of Physics, Zhejiang University, Hangzhou 310058, China}
 \affiliation{Institute for Advanced Study in Physics, Zhejiang University, Hangzhou 310058, China}

\author{Hai-Qing Lin}
 \email[E-mail address: ]{hqlin@zju.edu.cn}
 \affiliation{School of Physics, Zhejiang University, Hangzhou 310058, China}
 \affiliation{Institute for Advanced Study in Physics, Zhejiang University, Hangzhou 310058, China}

\date{\today}

\begin{abstract}
The \ccs\ exhibits superconductivity in close proximity to a density-wave (DW) like ground state at ambient pressure\cite{Liu:2024aa}, however details of the DW is still elusive. Using first-principles density-functional calculations, we found its ground state to be a $4\times2$ altermagnetic spin-density-wave (SDW) at ambient pressure, with an averaged effective moment of $\sim$1.7$\mu_B$/Cr. The magnetic long range order is coupled to the lattice, generating 4$a_0$ structural modulation. Multiple competing SDW phases are present and energetically close, suggesting strong magnetic fluctuation at finite temperature. The electronic states near Fermi level are dominated by Cr-3$d$ orbitals, and the kagome flat bands are closer to the Fermi level than those in the $A$V$_3$Sb$_5$ family in paramagnetic state. When external pressure is applied, the energy differences between competing orders and structural modulations are suppressed. Yet, the magnetic fluctuation remains present and important even at high pressure because the high-symmetry kagome lattice is unstable in nonmagnetic phase up to 30 GPa. Our results suggest the crucial role of magnetism to stabilize the crystal structure, under both ambient and high pressure.  

%\hl{Our findings strongly support the robust magnetic ground state of} \ccs\ \hl{ compound and suggest magnetic fluctuations might potentially lead to the unconventional superconductivity in the}  \ccs \hl{after the structure transition and magnetic long-range order are suppressed by pressure.}

%Our findings strongly support unconventional superconductivity in the \ccs\ compound above 5 GPa, and suggest crucial role of magnetic fluctuations in the pairing mechanism.
\end{abstract}

%\pacs{74.20.Pq, 74.70.Dd, 75.70.Tj}
\maketitle

%\newpage
%\thispagestyle{empty}

%\newpage

\section{Introduction}
Kagome lattice has been an intriguing system\cite{yinTopologicalKagomeMagnets2022} since its original proposal in 1951\cite{FirstKagome}. As one of the most frustrated geometry, the ground state of spin-1/2 kagome lattice is widely believed to be quantum spin liquid (QSL), but its nature is still under hot debate\cite{yanSpinLiquidGroundState2011,messioKagomeAntiferromagnetChiral2012,depenbrockNatureSpinLiquidGround2012,xieTensorRenormalizationQuantum2014,gongGlobalPhaseDiagram2015,bieriGaplessChiralSpin2015,liaoGaplessSpinLiquidGround2017}. Once slightly doped, a superconducting phase may emerge from the proposed $U(1)$ Dirac QSL\cite{jiangPossibleSuperconductivityBogoliubov2021}. For spin-1 kagome lattices, model studies point to trimerized or simplex valence bond solid as its ground state\cite{caiSpontaneousTrimerizationTwodimensional2009,changlaniTrimerizedGroundState2015,liuSimplexValencebondCrystal2015}. Beyond the spin models, the electronic and phonon band structure of kagome lattice exhibits topological flat bands due to the destructive interference between Bloch waves, in addition to the Dirac cones and van Hove singularities (vHS)\cite{PhysRevLett.106.236802,PhysRevLett.121.096401}. Notably, nontrivial physical properties including kinetic ferromagnetism\cite{pollmannKineticFerromagnetismKagome2008}, fractional Quantum Hall effect\cite{PhysRevLett.106.236802} and large band-specific diamagnetism\cite{NegativeFlatBand} may emerge from the flat bands. Moreover, near the van Hove fillings, different orders include charge density wave (CDW), chiral spin density wave (cSDW), chiral $d$-wave superconductivity or even $f$-wave triplet superconductivity are proposed\cite{PhysRevB.79.214502,yuChiralSuperconductingPhase2012,PhysRevLett.110.126405,wangCompetingElectronicOrders2013a}.

Of the many kagome materials, the ternary \avs\ family has drawn much attention since its discovery\cite{PhysRevMaterials.3.094407}. Their much debated CDW order\cite{PhysRevLett.125.247002,PhysRevX.11.031050,PhysRevB.104.075148,PhysRevLett.127.217601,PhysRevLett.127.187004,Zhao:2021aa,Li:2022aa,Chen:2021aa} is widely believed to be associated with the nesting between vHSs\cite{PhysRevX.11.041010,Kang:2022aa}, and may be chiral and break time-reversal symmetry (TRS)\cite{Jiang:2021aa,FENG20211384,yu2021evidence,PhysRevB.104.035131,PhysRevB.104.075148,Mielke:2022aa}. The superconductivity was proposed to be unconventional\cite{PhysRevLett.127.177001,zhao2021nodal,Chen:2021aa}, but evidence for conventional BCS superconductivity are also presented\cite{PhysRevB.103.L241117,Duan:2021aa,Mu_2021,PhysRevB.108.L060503,Zhong:2023aa}. Under pressure, the CDW is suppressed and the superconductivity exhibits two-dome structure\cite{PhysRevB.103.L220504,PhysRevB.105.094507}. Nevertheless, despite of its rich physics and phenomena, it is widely accepted that \avs\ family is weakly correlated without intrinsic magnetism\cite{PhysRevB.103.L241117,Kenney_2021}.

Recently, a chromium-based kagome compound \ccs\ was reported to be superconducting at pressure $p>p_c=4$ GPa\cite{Liu:2024aa}.  Under ambient pressure, the \ccs\ compound crystallizes in hexagonal structure with space group P6/mmm (No. 191) at room temperature, similar to the \avs\ family.  Below 55K, a stripe-like $4a_0$ structural modulation occurs, possibly due to CDW, with a concurrent SDW phase transition. External pressure continuously suppresses the long range magnetic order as well as the CDW modulation, which eventually disappear at around 4GPa where superconductivity appears with highest $T^{\mathrm{max}}_c=6.4$ K at $p_m=$4.2 GPa. Beyond $p_m$, the $T_c$ decreases, forming a dome-like structure. The upper critical field $H_{c2}$ is well beyond the Pauli limit around the $T^{\mathrm{max}}_c$, suggesting unconventional pairing mechanism. Several imminent questions are therefore needs to be addressed: 1) What is the long-range AFM pattern in this highly frustrated lattice?  2) What is the nature of the ordered magnetism? In particular, is the long range order local moment or itinerant? 3) How does the electronic structure evolve under pressure?

In this paper, we perform systematic study on \ccs\ using first-principles simulations, focusing on its ground state at ambient pressure and the effect of external pressure on the electronic structure. The ground state at ambient pressure is a collinear $4\times2$ altermagnetic spin-density-wave (SDW) order. In addition, multiple SDW orders are energetically close to each other, suggesting strong fluctuation at finite temperature. Under pressure, the competition among different magnetic phases enhances, and a Lifshitz transition due to metallization of a bonding state is observed at the critical pressure $p_c$. Distinguished from \avs, the magnetism is crucial to stabilize the lattice structure even at high pressures.

\section{Results and Discussion}

\subsection{High Symmetry Phase under Ambient Pressure}
The high-temperature structure of \ccs\ without lattice distortion is isostructural to \avs, where the Cr atoms are arranged into corner-sharing triangles, forming the kagome lattice (FIG. \ref{fig:crystal}a-b). Under ambient pressure, the fully optimized structure using nonmagnetic (NM) DFT calculations yields lattice constant $a$ and all bond lengths smaller than experimental value (refer to SI for detail). This is in sharp contrast to the \avs, whose lattice constants and bond lengths are slightly overestimated in NM DFT calculations due to the underestimate of bonding in general gradient approximation (GGA). Such phenomenon, however, is frequently observed in compounds where magnetic fluctuations are important, including iron-pnictides/chalcogenides\cite{PhysRevB.77.220506,PhysRevLett.100.237003,PhysRevB.78.085104} and other chromium based superconductors\cite{Jiang:2015aa,Wu_2015}. 

The electronic structure of the high-symmetry phase \ccs\ at high temperatures is shown in FIG. \ref{fig:bs_dos}a. For undistorted NM \ccs, each Cr atom is surrounded by 4 Sb$^{\mathrm{II}}$ atoms located directly above/below the center of kagome triangles, as well as 2 Sb$^{\mathrm{I}}$ atoms located at the center of kagome hexagons. Most of the electronic bands are dominated by the Cr-3$d$ orbitals, except for one dispersive band from $p_z$ orbitals of Sb and hybridizes with $d_{yz}$/$d_{xz}$ orbitals along $\Gamma$-M and $\Gamma$-K. Flat bands formed by $d_{xz}$/$d_{yz}$ orbitals can be identified around 300 meV above the Fermi level, which are considerably closer to the Fermi level compared to those in the $A$V$_3$Sb$_5$ family\cite{PhysRevLett.127.046401}. Other kagome band characteristics, including the vHSs and Dirac cones can still be identified from the large-scale plots, but they are mostly far from the Fermi level. Interestingly, additional vHS can also be identified around K, most prominently around 270 meV below $E_F$. Along K-M, this band exhibits rather flat dispersion, giving rise to an unambiguous vHS in the DOS (Fig.~\ref{fig:bs_dos}a). Density of state and symmetry analysis suggest it is a bonding state between Cr and Sb$^{\mathrm{II}}$ atoms. Another nearly flat band due to the $d_{z^2}$-orbitals can also be identified around K, but no vHS can be identified from the DOS. The bands are also very flat along $\Gamma$-A, implying the quasi-two-dimensionality of its electronic structure. The NM phonon spectrum of \ccs\ under ambient pressure using high symmetry kagome structure exhibits enormous imaginary frequencies at almost every $\mathbf{q}$-point along the high symmetry lines (FIG. \ref{fig:phonon}a,b), suggesting the high-symmetry lattice is highly unstable if the system is purely nonmagnetic. 

\subsection{Lattice Distortion and Magnetism under Ambient Pressure}

\begin{figure}
\includegraphics[width=8.5cm]{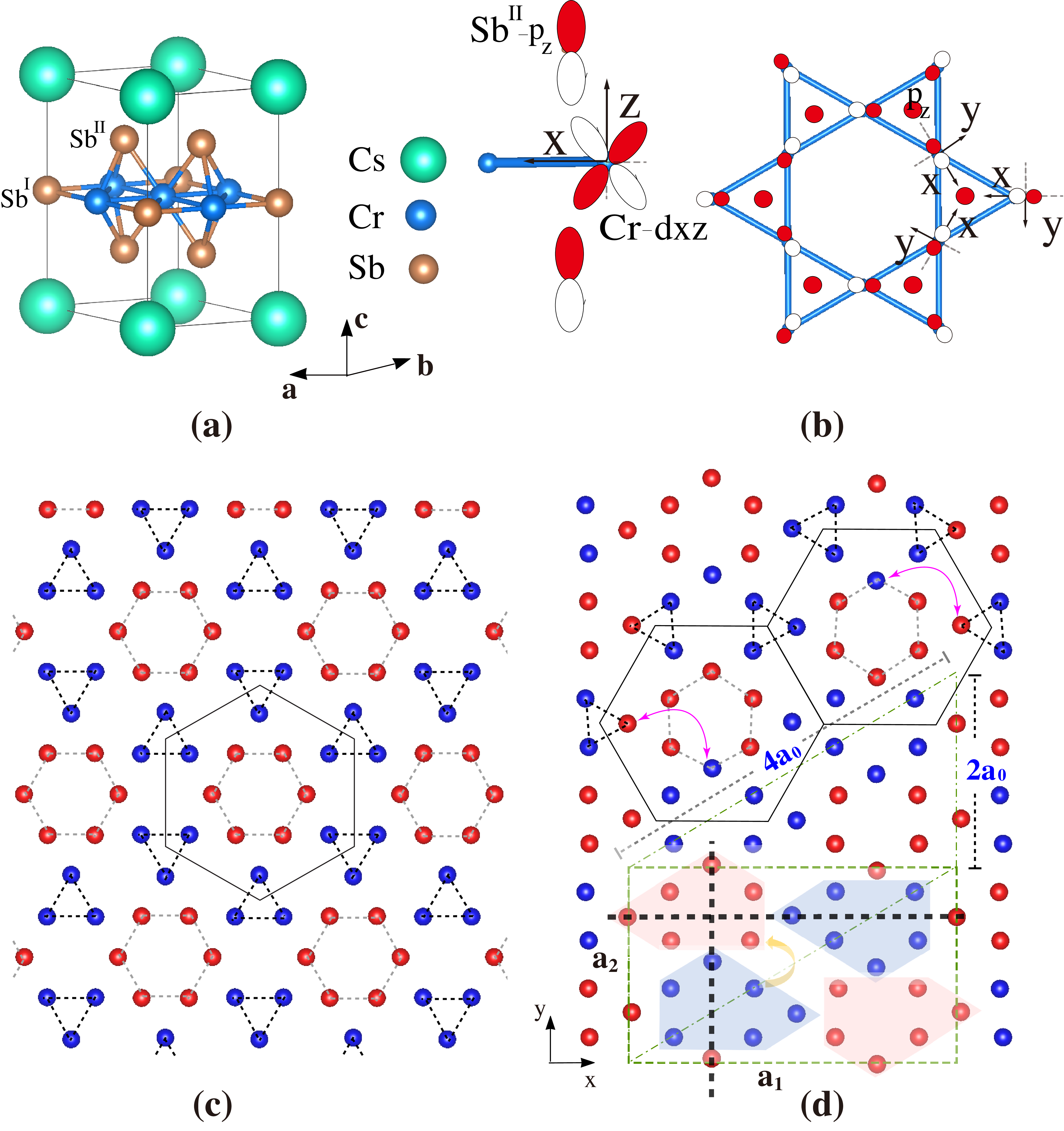}
 \caption{\textbf{High-temperature crystal structure and low-temperature SDW pattern.}
 (a) Crystal structure of \ccs\ with P6/mmm symmetry. (b) Definition of atomic orbitals of Cr atoms. (c) AF-SOD SDW pattern inspired by inverse star-of-david charge order. The black solid hexagon indicates its Wigner-Seitz (WS) unit cell. (d) Ground state SDW pattern (failed AF-SOD) of \ccs\ at ambient pressure. The black solid hexagons and pink arrows indicate the original WS unit cells and the swapped moments compared to the AF-SOD pattern. The thick black dashed lines indicate the gliding-mirror planes. The dot-dashed line shows the $4\times2$ supercell. The blue/red shaded area helps to identify the symmetry between two spin-sublattices. In (c-d), the spin-up and spin-down sublattices are marked in red and blue, respectively.  
}
 \label{fig:crystal}
\end{figure}

\begin{figure*}[thp]
\includegraphics[width=16cm]{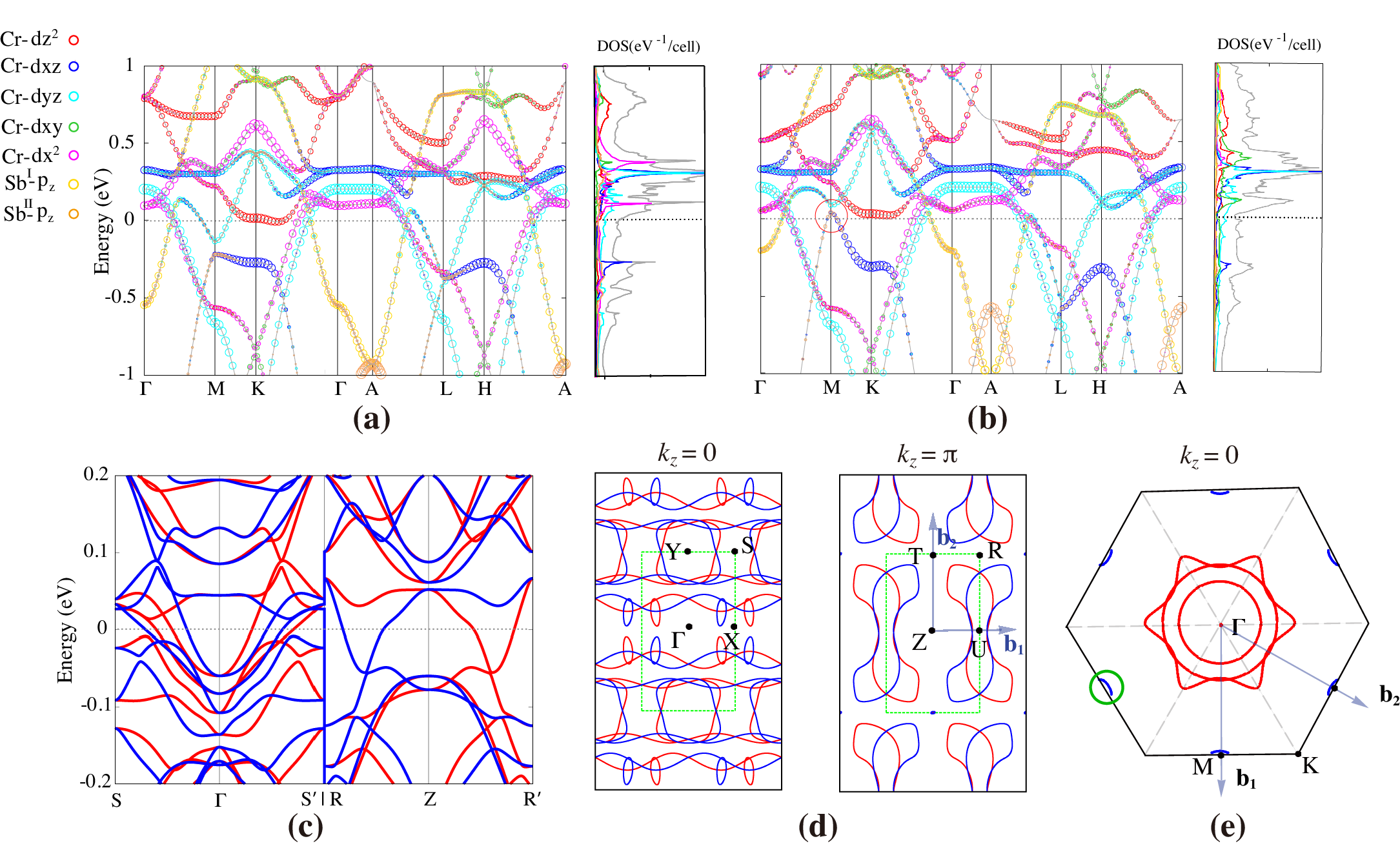}
 \caption{\textbf{Electronic structures and Fermi surface of NM state and failed AF-SOD ground state.}
 (a, b) Band structure and density of states (DOS) of \ccs\ in the high-symmetry NM state at (a) ambient pressure and (b) 5 GPa. The size of circles are proportional to the weight of corresponding atomic orbitals. The red circle is used to emphasize the most significant pressure effect on the NM band structure. The Cr-3d atomic orbitals are defined in FIG. \ref{fig:crystal}b. (c) Band structure of \ccs\ in the failed AF-SOD ground state at ambient pressure. (d) Intersection of failed AF-SOD ground state Fermi surfaces at $k_z=0$ and $k_z=\pi$. The First Brillouin zone is illustrated with green solid line.  For (c) and (d), the red/blue lines correspond to spin-up/down, respectively; and the high symmetry points are chosen according to the standardized unit cell. (e) The intersection of NM Fermi surfaces at $k_z=0$ plane at 5 GPa. The additional Fermi surface pocket around M due to the Lifshitz transition is indicated by green circle.}
\label{fig:bs_dos}
\end{figure*}

\begin{figure*}[htb]
\includegraphics[width=16cm]{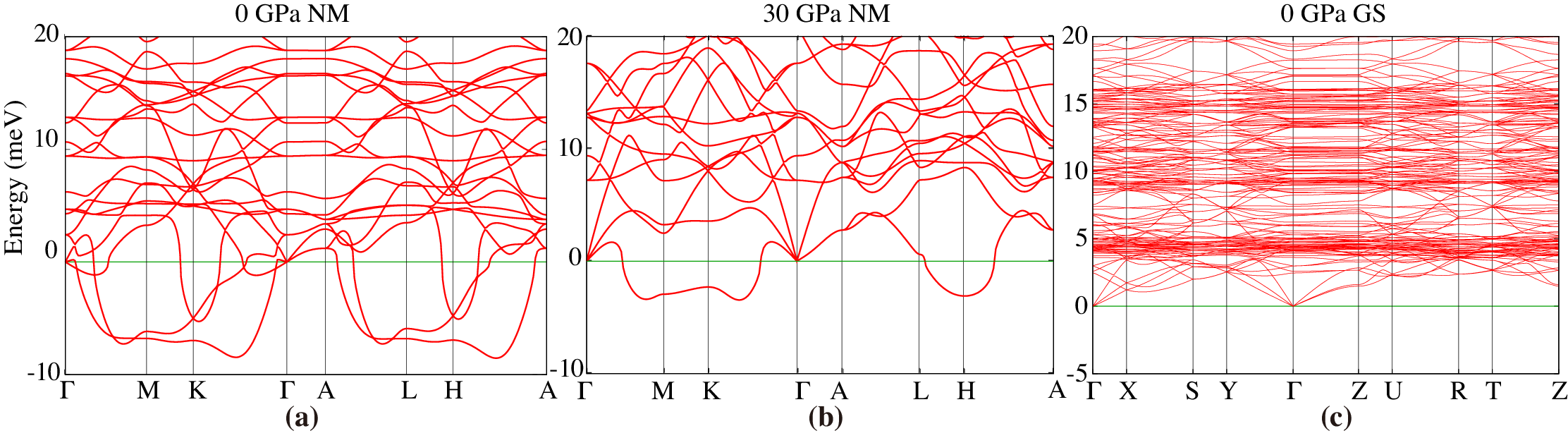}
 \caption{\textbf{Phonon spectrum of NM state and failed AF-SOD ground state.}
 (a, b) The phonon spectrum of \ccs\ in high-symmetry NM state at (a) ambient pressure and (b) 30 GPa. (c) The phonon spectrum of \ccs\ in failed AF-SOD ground state at ambient pressure. The corresponding magnetic pattern is illustrated in FIG.~\ref{fig:crystal}d. The zero frequency is marked with green solid line.} 
\label{fig:phonon}
\end{figure*}

In order to figure out the magnetic ground state of \ccs\ under ambient pressure, we first sort out several candidates proposed in previous model studies of spin-1/2 and spin-1 kagome lattices. In particular, collinear patterns including FM, up-up-down (UUD), A-type AFM, as well as noncollinear patterns including $1\times1$ 120$^{\circ}$-AFM, $\sqrt{3}\times\sqrt{3}$ in-plane AFM I/II, $2\times2$ cuboctahedron ({\it cuboc}) phases are considered. Inspired by the star-of-David and inverse star-of-David (SOD/ISOD) charge orders \cite{PhysRevLett.127.046401}, we have also considered  2 $2\times2$ antiferromagnetic SOD patterns, dubbed as in-out SOD and all-in (or all-out) SOD patterns (refer to SI). In addition, a collinear equivalent of such pattern, namely AF-SOD (FIG. \ref{fig:crystal}c), is also considered. In the AF-SOD pattern, all Cr atoms are coordinated with exactly 2 nearest antiparallel Cr atoms, such that one can arrange nearest neighboring AF bonds to form SOD pattern. It is important here to point out that the AF-SOD pattern is a ferrimagnetic state, because the spin-up sublattice is inequivalent to the spin-down sublattice and there is no symmetry to guarantee overall 0 net moment. In fact, there is a residue overall total moment of  1.36 $\mu_B$/cell for the AF-SOD phase in calculation.

Our initial calculations show that, the collinear orders prevail the noncollinear orders at the same magnetic cell size (TAB. \ref{tab:mag}). In addition, patterns with nonzero net moment are energetically less favorable, indicating overall AFM tendency.  Therefore, we employed a high-throughput algorithm (please refer to METHODS for detail) to search for the collinear AFM configurations with lowest energy. Considering the structural modulation with a single $\mathbf{Q}$ vector (1/4, 0, 0) observed in experiment\cite{Liu:2024aa}, all $2\times1$, $2\times2$, $4\times1$, $8\times1$ and $4\times2$ collinear AFM orders are searched without considering spin-orbit coupling (SOC). Overall, the lattice constants and bond lengths of fully relaxed structures in most magnetic phases are closer to experimental observations. The averaged magnetic moment is quite robust around 1.7 $\sim$ 1.8 $\mu_B$/Cr in all calculations even without explicitly considering the on-site Coulomb interactions, except for the in-plane FM cases (1.5 $\mu_B$/Cr).

%Therefore, we performed a complete search of all possible collinear AFM orders up to 8 primitive cells with 24 magnetic sites. To do this, we first generate all symmetrically inequivalent configurations with 0 net moment, and then eliminate configurations without symmetry connecting the two sublattices. This procedure yields 3 $2\times1$, 5 $2\times2$, 15 $4\times1$, 760 $8\times1$ and 1128 $4\times2$ configurations. Employing a systematic high-throughput algorithm (refer to SI for detail), we search for the magnetic ground state for \ccs. Overall, 

The lowest energy magnetic configuration turns out to be a complex pattern (dubbed as ``failed AF-SOD'', shown in FIG. \ref{fig:crystal}d). It can be regarded as a decendent of the AF-SOD phase by swapping two pairs of the magnetic moment of next nearest neighboring atoms inside two neighboring $2\times2$ AF-SOD unit cells. The swapped atoms are uniformly distributed across all SOD structures and the staggered magnetic moment ranges from 1.4 to 1.9 $\mu_B$ for each Cr atom, aligning in parallel to minimize the total energy. We have also performed additional calculations with spin-orbit coupling (SOC) included, and the ``failed AF-SOD" pattern remains the lowest energy phase. The SOC-included calculations also indicate magnetic anisotropy is small, as the in-plane alignment of moments is only $\sim$0.6 meV/Cr lower than the out-of-plane alignment, and the in-plane anisotropy appears to be negligible(please refer to SI for details). In such situation, it is recently argued that the spin-space group is more appropriate to describe the symmetry of the phase\cite{PhysRevX.12.021016,PhysRevX.12.031042,PhysRevX.14.031038}.  The corresponding spin-space group is $P^{-1} b^{-1} a^1 m^{\infty m} 1$ (please refer to SI for details). It is evident that the two spin-sublattices are not connected spatially by either simple translation or inversion, but by $\lbrace M_{yz}|(0, 1/2, 0)\rbrace$ or $\lbrace M_{xz}|(1/2, 0, 0)\rbrace$, a gliding-mirror operation purely on the spatial part. Thus, the ground state at the ambient pressure is altermagnetic\cite{PhysRevX.12.040501}. We note that such altermagnetic ground state is robust under DFT+$U$ calculations with abinitio $U$ and $J$ values determined from constrained random phase approximation (cRPA) calculations (please refer to SI for details).

The formation of magnetic long range order has nontrivial effect on the lattice structure by introducing structural modulations. In our calculations, all lowest energy configurations breaks the 3-fold rotational symmetry, and both the Cr and the Sb$^{\mathrm{II}}$ positions are strongly modulated. In particular, the failed AF-SOD ground state is formed by 2 $4\times1$ stripes connected by gliding-mirror operation. Similarly, the second lowest (indexed with 6707797) naturally consists of 2 $4\times1$ stripe order connected by translation only, which is also consistent with experimentally observed $4\times1$ charge order at low temperature. The calculated phonon spectrum of the failed AF-SOD phase is free of imaginary phonon frequency (FIG.~\ref{fig:phonon}c), indicating the dynamic stability of this phase. However, the phonon spectrum of the corresponding fully-relaxed non-magnetic structure is not (FIG.~S3). Thus, the magnetism stabilizes the lattice structure in \ccs. 

%We illustrate the electronic structure of \ccs\ at the ambient pressure in FIG. \ref{fig:bs_dos}a. For undistorted NM \ccs, each Cr atom is surrounded by 4 Sb$^{\mathrm{II}}$ atoms located directly above/below the center of kagome triangles, as well as 2 Sb$^{\mathrm{I}}$ atoms located at the center of kagome hexagons. Most of the electronic bands were dominated by the Cr-3d orbitals, except for one dispersive band from p$_z$ orbitals of {\color{red} \st{Sb}$^{\mathrm{II}}$} Sb and hybridizes with d$_{yz}$/d$_{xz}$ orbitals along $\Gamma$-M and $\Gamma$-K. Flat bands formed by d$_{xz}$/$d_{yz}$ orbitals can be identified around 300 meV above the Fermi level, which are considerably closer to the Fermi level compared to those in the $A$V$_3$Sb$_5$ family\cite{PhysRevLett.127.046401}. Other kagome band characteristics, including the vHSs and Dirac cones can still be identified from the large-scale plots, but they are mostly far from the Fermi level. Interestingly, additional vHS can also be identified around K, most prominently around 270 meV below $E_F$. Along K-M, this band exhibits rather flat dispersion, giving rise to the unambiguous vHS in the DOS (Fig.~\ref{fig:bs_dos}a). Density of state and symmetry analysis suggest it is a bonding state between Cr and Sb$^{\mathrm{II}}$ atoms. Another nearly flat band due to the d$_{z^2}$-orbitals can also be identified around K, but no vHS can be identified from the DOS. The bands are also very flat along $\Gamma$-A, implying the quasi-two-dimensionality of its electronic structure. 

%For \ccs\ in the failed AF-ISOD ground state, 
The spin-degeneracy of the failed AF-SOD ground state is generally lifted (FIG. \ref{fig:bs_dos}c) due to the lack of inversion or simple translation symmetry between sublattices (FIG. \ref{fig:crystal}d).  Although the quasi-Kramer's degeneracy protected by SSG symmetries can still be identified along certain high symmetry lines\cite{PhysRevX.12.021016}, the spin splitting is evident along $\Gamma$-S, with the largest splitting $\sim$80 meV close to the Fermi level. The splitting suggests separated spin-up/spin-down Fermi surface sheets (FIG. \ref{fig:bs_dos}d, refer to SI for more details). These features may be verified through spin-resolved ARPES measurements to confirm the altermagnetic ground state\cite{Krempasky2024-dj,PhysRevLett.132.036702}, if the measurement can be constrained within a single domain. In addition, our calculations show that moderate anomalous Hall conductivity may be present if the direction of magnetic moments are in-plane and the field direction is properly aligned (refer to SI for detail)\cite{PhysRevX.12.011028,Smejkal2022-wi}. It is worth noting that the flat-bands and vHS features are less obvious in the magnetic ground state compared to the undistorted NM \ccs, due to the strong structure distortion of the altermagnetic ground state (refer to SI for detail).

 To understand the origin of the magnetism, we have tried to fit the DFT total energies of 20 lowest magnetic configurations to a Heisenberg spin model with the bilinear exchange interactions. However, such fitting is not successful up to 6$^{th}$-nearest-neighboring exchange interactions ($\sim$ 9.3 \AA\ apart, please refer to SI for details). Considering the moments on Cr atoms have a wide distribution in all collinear magnetic configurations, our results suggest that the Fermi-surface-related itinerant magnetism may play an important role in the formation of the SDW order.

\subsection{Pressure Effect}
With the knowledge at ambient pressure, we now investigate the pressure effect. The total energy, geometry and Cr magnetic moment of the leading magnetic instabilities under different pressure are also shown in TAB. \ref{tab:mag}. Under pressure, the magnitude of magnetic moment on each Cr atom $m_{\mathrm{Cr}}$ are only slightly affected, but the total energy difference between the competing magnetic phases $\Delta E$ are significantly reduced. In particular, $\Delta E$ between the ground state and lowest energy competing phases of $2\times2$, $4\times2$, $4\times1$ and $8\times1$ is reduced to less than 1 meV/f.u..  We argue that the long range magnetic order would be suppressed at high pressure, if the quantum fluctuations are fully considered. Nevertheless, the magnetism remains important even at high pressure, as the phonon calculations indicate the structure instability is robust under pressure up to 30 GPa in NM state (FIG. \ref{fig:phonon}b). Experimentally, however, the structural transition occurs at relatively low temperature ($\sim$55K) at ambient pressure, and is quickly suppressed by applying external pressure at 5 GPa. Therefore, we conclude that the electronic correlations and magnetic fluctuations are crucial in the current system of interest, even at high pressure/temperatures that long range magnetic orderings are suppressed. As a result, if the static altermagnetic long-range order is continuously suppressed, superconductivity with exotic pairing symmetry is possible due to the dynamic altermagnetic fluctuations \cite{PhysRevX.12.040501,mazin2022notes,Zhang:2024aa}. Furthermore, the external pressure has nontrivial effect on the electronic structure as well. Most remarkably, the bonding state originally forming the vHS around K becomes dispersive along M-K under pressure(red circle in FIG. \ref{fig:bs_dos}b), and the vHS gradually vanishes. Around 5 GPa, a Lifshitz transition occurs and the aforementioned bonding state crosses the Fermi level (green circle in FIG. \ref{fig:bs_dos}e). Such metallization of bonding state may also affect the superconductivity\cite{PhysRevB.104.L100504,Sun:2023aa}. 

In conclusion, we have performed systematic study of the electronic structure, magnetism and lattice stability of \ccs. The ground state of \ccs\ at the ambient pressure is found to be $4\times2$ collinear altermagnetic SDW-type order, with considerable itinerant magnetism component. Under high pressure, the energy differences between competing orders are significantly suppressed, suggesting enhanced magnetic fluctuations. In addition, the NM phonon calculations indicate the high-symmetry kagome structure is unstable at low temperature up to 30 GPa. Similarly, the NM phonon spectrum is also highly unstable even for the distorted structure of the ground state. In comparison, the distorted structure becomes stable after the magnetism is considered. These results suggest the magnetic fluctuation is extremely important and may couple to the lattice dynamics in \ccs. Therefore, a simple electron-phonon coupling based BCS mechanism is highly unlikely in this compound.

\section{Methods}
\textbf{Electronic and phonon band structure} The calculations were performed based on density functional theory (DFT)  with \textsc{Vienna Abinitio Simulation Package} (VASP)\cite{VASP_Kresse_PRB93,VASP_Kresse_PRB99} and cross-checked with Quantum \textsc{Espresso} (QE)\cite{Qe}.  The energy cutoff of plane-wave basis was 450 eV and a $\Gamma$-centered 12$\times$12$\times$6 $\mathbf{k}$-point mesh were employed in the self-consistent calculations with VASP. QE calculations were performed with ultrasoft-pseudopotentials with energy cutoff 64 Ry and 720 Ry for the wavefunction and augmentation charge, respectively. In all calculations, the PBEsol approximation \cite{PBEsol} was used, and spin-orbit coupling (SOC) was not included unless otherwise stated. The lattice constants and atomic coordinates were fully relaxed until the force on each atom was less than 1 meV/\AA\  and internal stress was less than 0.1 kbar. After full structure optimization, the phonon calculations were also performed with 4$\times$4$\times$3 $\mathbf{q}$-point mesh for high-symmetry NM state using density-functional perturbation theory (DFPT) as implemented in QE. For the phonon spectrum of the failed AF-SOD state, finite displacement method using $1\times2\times2$ supercell is employed, as implemented in \textsc{PHONOPY}\cite{phonopy-phono3py-JPCM}. 

\textbf{High-throughput Search for magnetic configuration.} The high-throughput calculations were performed for all possible collinear magetic patterns up to $4\times2$ supercell.
 \begin{enumerate}
   \item{ First, all possible magnetic configurations are indexed by assigning a $N$-bit binary integer to each configuration, where $N$ is the number of the Cr sites within the supercell. For example, $7=\underline{000111}_2$ in $2\times1$ configuration means the first 3 sites are spin-down and last 3 sites spin-up. The configurations with none-zero net moment are then discarded. Then, we search for the AFM symmetry connecting up/down-sublattices within each configuration, and eliminate those without such symmetry. In addition, only 1 of all symmetrically equivalent configurations are kept. After this procedure, there are in total 1911 configurations. The $8\times1$ and $4\times2$ supercells generate 1888 configurations, whereas the rest 3 generates 23 configurations.}
   \item{ Secondly, for $2\times1$, $2\times2$ and $4\times1$ configurations, both static total energy calculations on perfect kagome lattice and full geometry relaxations were performed. Therefore, for each configuration, we obtain both $E_0$ and $E_r$. The former represents magnetic configuration energy difference only, while the latter is actual total energy. We note that for all these configurations, the full relaxation yields energy difference order of 200 meV.}
   \item{ Thirdly, for $4\times2$ and $8\times1$ configurations, a static total energy calculation on perfect kagome lattice is performed first. Then, the configurations with $E_0$ over 200 meV higher than the lowest $E_0$ are screened out. The remaining configurations are fully relaxed to obtain $E_r$.}
   \item{ Finally, the configurations with lowest $E_r$ are verified with dense K-mesh and SOC calculations.}
 \end{enumerate} 
 
\textbf{Anomalous Hall conductivity.}  For the anomalous Hall conductivity calculations, we include SOC into the calculations.  The DFT results were then fitted to a tight-binding (TB) Hamiltonian with maximally projected Wannier function method \cite{Wannier90}.  30 atomic orbitals including Cr-$3d$ and Sb-$5p$  were chosen to construct the TB model Hamiltonian.  The resulting Hamiltonian was symmetrized using WannSymm code\cite{ZHI2022108196}. With the symmetrized TB Hamiltonian, the intrinsic anomalous Hall conductivity (AHC) is calculated following Kubo formula by integration of Berry curvature over the Brillouin zone\cite{Wang_ahc,Yao_ahc}:

\begin{equation}
\begin{aligned}
& \Omega_n^{\alpha \beta}(\mathbf{k})=\frac{2 e^2}{\hbar} \sum_{m \neq n} \frac{\operatorname{Im}\left[\left\langle n\left|v_\alpha\right| m\right\rangle\left\langle m\left|v_\beta\right| n\right\rangle\right]}{\left(E_n-E_m\right)^2} 
\end{aligned}
\end{equation}
\begin{equation}
\begin{aligned}
\sigma_{\alpha \beta}=-\frac{e^2}{\hbar} \sum_{\mathbf{k} n} \frac{d^3 \mathbf{k}}{(2 \pi)^3} f_{\mathbf{k} n} \Omega_n^{\alpha \beta}(\mathbf{k})
\end{aligned}
\end{equation}

The implementation in WannierTools\cite{WU2017} were employed to calculated the AHC. To fix collinear moments in [100], [010] or [001] directions, SOC were applied. The Berry curvature was calculated with a $100\times100\times100$ $\Gamma$-centered kmesh, with which the convergence of $\sigma_{\alpha \beta}$ is achieved. 
 
 \textbf{Constrained random-phase calculation.}  We performed the constrained random-phase approximation (cRPA) method using VASP to calculate the effective interaction parameter $U$ and $J$. With the 30 atomic-like Wannier orbitals, we constructed the target space with Cr-$3d$ orbitals:
\begin{equation}
\left|w_i^\sigma\right\rangle=\frac{1}{N_k} \sum_{n \mathbf{k}} T_{i n}^{\sigma(\mathbf{k})}\left|\psi_{n \mathbf{k}}^\sigma\right\rangle 
\end{equation}
The  Coulomb interaction matrix elements are written as:
\begin{equation}
 \begin{split}
U_{i j k l}^{\sigma \sigma^{\prime}}=\frac{1}{N_k^3} \sum_{\mathbf{k k' q}} \sum_{n_1 n_2 n_3 n_4} T_{i n_1}^{*(\mathbf{k})} T_{j n_2}^{(\mathbf{k}-\mathbf{q})}\left\langle u_{n_1 \mathbf{k}}\left|e^{-i(\mathbf{q}+\mathbf{G}) \cdot \mathbf{r}}\right| u_{n_2 \mathbf{k}-\mathbf{q}}\right\rangle U_{\mathbf{G} \mathbf{G}^{\prime}}(\mathbf{q}) \\ 
\times \left\langle u_{n_3 \mathbf{k}^{\prime}-\mathbf{q}}\left|e^{i\left(\mathbf{q}-\mathbf{G}^{\prime}\right) \cdot \mathbf{r}^{\prime}}\right| u_{n_4 \mathbf{k}^{\prime}}\right\rangle T_{k n_3}^{*\left(\mathbf{k}^{\prime}-\mathbf{q}\right)} T_{l n_4}^{\left(\mathbf{k}^{\prime}\right)} 
\end{split}
\end{equation}
where the effective Coulomb kernel $U_{\mathbf{G}\mathbf{G'}}(\mathbf{q})$ can be calculated as
 \begin{equation}
U_{\mathbf{G G}^{\prime}}^{\sigma \sigma^{\prime}}(\mathbf{q}, i \omega)=\left[\delta_{\mathbf{G G}^{\prime}}-\left(\chi_{\mathbf{G G}^{\prime}}^{\sigma \sigma^{\prime}}(\mathbf{q}, i \omega)-\tilde{\chi}_{\mathbf{G G}^{\prime}}^{\sigma \sigma^{\prime}}(\mathbf{q}, i \omega)\right) \cdot V_{\mathbf{G G}^{\prime}}(\mathbf{q})\right]^{-1} V_{\mathbf{G G}^{\prime}}(\mathbf{q}) 
\end{equation}
$\tilde{\chi}_{\mathbf{G G}^{\prime}}^{\sigma \sigma^{\prime}}(\mathbf{q}, i \omega)$ denotes the polarizability within the target space. The screening effect due to the target polarizability is excluded from the total screening effects in RPA method. Using projector method\cite{10.25365/thesis.38099}, the target polarizability is obtained with :
 \begin{equation}
 \begin{split}
\tilde{\chi}_{\mathbf{G}, \mathbf{G}^{\prime}}^\sigma(\mathbf{q}, i \omega)=\frac{1}{N_k} \sum_{n n^{\prime} \mathbf{k}} \frac{f_{n \mathbf{k}}-f_{n^{\prime} \mathbf{k}-\mathbf{q}}}{\epsilon_{n \mathbf{k}}-\epsilon_{n^{\prime} \mathbf{k}-\mathbf{q}}-i \omega} \\ \times 
\sum_{m_1 m_2^{\prime}} P_{m_1 n}^{* \sigma(\mathbf{k})}\left\langle u_{m_1 \mathbf{k}}^\sigma\left|e^{-i(\mathbf{G}+\mathbf{q}) \mathbf{r}}\right| u_{m_2^{\prime} \mathbf{k}-\mathbf{q}}^{\sigma^{\prime}}\right\rangle P_{m_2^{\prime} n^{\prime}}^{\sigma^{\prime}(\mathbf{k}-\mathbf{q})} \\\times \sum_{m_1^{\prime} m_2} P_{m_2 n^{\prime}}^{* \sigma^{\prime}(\mathbf{k}-\mathbf{q})}\left\langle u_{m_2 \mathbf{k}-\mathbf{q}}^{\sigma^{\prime}}\left|e^{-i\left(\mathbf{G}^{\prime}-\mathbf{q}\right) \mathbf{r}^{\prime}}\right| u_{m_1^{\prime} \mathbf{k}}^\sigma\right\rangle P_{m_1^{\prime} n}^{\sigma(\mathbf{k})}  
\end{split}
\end{equation}
where the the target projectors $P_{m n}^{\sigma(\mathbf{k})}$ for Cr-$3d$ orbitals are defined as
\begin{equation}
P_{m n}^{\sigma(\mathbf{k})}=\sum_{i \in \mathcal{T}} T_{i m}^{* \sigma(\mathbf{k})} T_{i n}^{\sigma(\mathbf{k})} 
\end{equation}
And the static limit ($\omega=0$) values are taken as the effective interaction parameters. The large bare interactions ($U^{\mathrm{bare}}_{iiii}\sim18$ eV, $U^{\mathrm{bare}}_{iijj}\sim17$ eV, $U^{\mathrm{bare}}_{ijij}\sim0.6$ eV) are strongly screened to yield $U=$1.0 eV and $J=$0.4 eV.

\section*{Data availability}
%The data that support the findings of this study are available from the corresponding author upon request.
The data used to generate FIG. \ref{fig:bs_dos} and \ref{fig:phonon} are available via Figshare at

https://doi.org/10.6084/m9.figshare.28595522 . The mcif files of the configurations in TAB. \ref{tab:mag} are also available as source data. Any additional data relevant to the findings of this study are available from the corresponding author upon request.

\newpage

\bibliographystyle{naturemag}
%bibliographystyle{apsrev4-1}
%\bibliography{135}

\section*{Acknowledgments}
The authors are grateful to Xiaofeng Xu, Jinke Bao, Yi Liu, Jiangfan Wang and Dexi Shao for stimulating discussions. C. C. acknowledges support from the National Key R\&D Program of China (Nos. 2024YFA1408303 \& 2022YFA1402202) and the National Natural Science Foundation of China (No. 12350710785 \& 12274364). X. W. and H.-Q. L. acknowledge support from the National Key R\&D Program of China (No. 2022YFA140271). G.-H. C. acknowledges the support from the Key R\&D Program of Zhejiang Province (2021C01002). C. X. acknowledges the support from the National Natural Science Foundation of China (No. 12304175). G.-X. Z. acknowledge the support from the Zhejiang Provincial Natural Science Foundation of China (LQ23A040014). The calculations were performed on clusters at the High Performance Computing Cluster at Center of Correlated Matters Zhejiang University and High Performance Computing Center at Hangzhou Normal University.

\section*{Author contributions}
C. C., G.-H. C.  and  C. X. initiated this work;  C. C., X. W. and H.-Q. L. supervised the research; C. X., S. W. and C. C. performed the calculations; C. X. and C. C. were responsible for the data analysis; C. X. and C. C. drafted the manuscript with input from all authors. C. X., S. W., G.-X. Z., G.-H. C., J. D., C. C., X. W. and H.-Q. L. participated in the discussion and revised the manuscript.

\section*{Competing Interests}
The authors declare no competing interests.

\begin{table}
\caption{ Total energies (in meV/f.u.) with respect to NM state and magnetic moment $m_{\mathrm{Cr}}$ (in $\mu_B$) of typical magnetic configurations. For AF-SOD state, the numbers in (outside) the parenthesis are the moments of Cr on the hexagons (triangles). The magnetic patterns can be found in SI. The SOC was included when comparing the energy between the collinear and noncollinear magnetic configurations. \label{tab:mag}}
  \begin{tabular}{c|c|c||cc|cc}  
     \hline
      \multirow{2}{*}{Size} & \multirow{2}{*}{Pattern} & \multirow{2}{*}{SG} & \multicolumn{2}{c|}{ 0 GPa}&  \multicolumn{2}{c}{ 5 GPa} \\
      \cline{4-7}
      &  &   & $E_{\mathrm{tot}}$ & $m_{\mathrm{Cr}}$ & $E_{\mathrm{tot}}$ & $m_{\mathrm{Cr}}$  \\
      \hline\hline
      \multirow{3}{*}{1$\times$1$\times$1}
      & FM    & P6/mmm   &-204.2  &  1.5 & -187.3  &  1.4 \\ 
      & UUD & P2/m   &-295.5 &  1.7 &  -215.4  & 1.3 \\ 
      & 120$^{\circ}$-AFM & P6/mmm   &-246.4 &  1.9 & \multicolumn{2}{c}{to NM} \\
     \hline
     1$\times$1$\times$2 
     & A-type AFM  &  P6/mmm & -200.8 &  1.5 &  -184.2  & 1.4 \\
     \hline
      %\multirow{2}{*}{$\sqrt{3}\times\sqrt{3}$ \\ $\times1$} 
      $\sqrt{3}\times\sqrt{3}$ &   AFM I & P$\bar{3}$1m & -297.6   &  1.8  & -238.6  &  1.6 \\   
      $\times1$ &  AFM II  &  P$\bar{3}$m1 &  -254.9   & 1.8  & -179.5  & 1.6 \\   
      \hline
      \multirow{6}{*}{2$\times$2$\times$1}
      & all-out SOD &  P6/mmm & -291.5  &  1.8  &  -223.2  &  1.6 \\
%      & all-out SOD & \multicolumn{4}{c}{converge to all-out SOD} \\
      & in-out SOD  &  P6/mmm & -228.9  &  1.8  &  -175.8.5 &  1.6 \\
%      & in-out SOD  & \multicolumn{4}{c}{converge to in-out SOD} \\
      & cuboc & \multicolumn{4}{c}{converge to 821}  \\
      & AF-SOD &   P6/mmm &-338.4   &  1.9 (1.7)  &  -275.8  &  1.8 (1.5) \\ 
      & 821  & Pm  & -382.8  &  1.7  &  -322.4.5  &  1.6 \\
      \hline
     4$\times$1$\times$1 &  
      915 &  Pmm2  & -387.3  &  1.8  &  -291.9   &  1.6 \\
      \hline
    \multirow{5}{*}{4$\times$2$\times$1} 
    & {\bf 6692437}\footnote{The line in bold font (6692437) is the failed AF-SOD ground state.} &    Pbam  & {\bf -404.0}  &  {\bf 1.7}  &  {\bf -329.7}  &  {\bf 1.6} \\ 
    & 6707797 & Pmm2  & -401.5  &  1.7  & -329.4  &  1.6 \\ 
    & 1550757 & P2/m  & -369.9   &  1.7  &  -323.5  &  1.6 \\ 
    & 5917545 & Cmmm  & -352.8   &  1.8  &  -296.4  &  1.6 \\ 
    & 4$\times$2 NCL &  \multicolumn{4}{c}{converge to 5917545}  \\
     \hline    
    8$\times$1$\times$1 &  2987565 & Pmm2 &-394.3 & 1.7  & -314.8 & 1.6\\
      \hline
  \end{tabular}
\end{table}
 \newpage

\end{document}